\documentclass[twoside,twocolumn,9pt]{article}
\usepackage{extsizes}
\usepackage[super,sort&compress,comma]{natbib} 
\usepackage[version=3]{mhchem}
\usepackage[left=1.5cm, right=1.5cm, top=1.785cm, bottom=2.0cm]{geometry}
\usepackage{balance}
\usepackage{mathptmx}
\usepackage{amsmath} 
\usepackage{amsfonts} 
\usepackage{amssymb} 
\usepackage{sectsty}
\usepackage{graphicx} 
\usepackage{lastpage}
\usepackage[format=plain,justification=justified,singlelinecheck=false,font={stretch=1.125,small,sf},labelfont=bf,labelsep=space]{caption}
\usepackage{float}
\usepackage{fancyhdr}
\usepackage{fnpos}
\usepackage[english]{babel}
\addto{\captionsenglish}{%
  
}
\usepackage{array}
\usepackage{droidsans}
\usepackage{charter}
\usepackage[T1]{fontenc}
\usepackage[usenames,dvipsnames]{xcolor}
\usepackage{setspace}
\usepackage[compact]{titlesec}
\usepackage{hyperref}
\hypersetup{colorlinks=true, linkcolor=blue, citecolor=blue, urlcolor=blue, pdftitle={Anionic nickel and nitrogen effects in the chiral antiferromagnetic antiperovskite Mn$_3$NiN}, pdfauthor={A. C. Garcia-Castro}}

\usepackage{orcidlink}

\usepackage{epstopdf}

\definecolor{cream}{RGB}{222,217,201}

\usepackage{comment} 
\usepackage{siunitx} 
\usepackage{amsmath,amsfonts,mathtools}
\usepackage{todonotes}
\usepackage[export]{adjustbox}

\begin{document}

\pagestyle{fancy}
\thispagestyle{plain}
\fancypagestyle{plain}{
\renewcommand{\headrulewidth}{0pt}
}

\makeFNbottom
\makeatletter
\renewcommand\LARGE{\@setfontsize\LARGE{17pt}{17}}
\renewcommand\Large{\@setfontsize\Large{12pt}{14}}
\renewcommand\large{\@setfontsize\large{10pt}{12}}
\renewcommand\footnotesize{\@setfontsize\footnotesize{7pt}{10}}
\makeatother

\renewcommand{\thefootnote}{\fnsymbol{footnote}}
\renewcommand\footnoterule{\vspace*{1pt}%
\color{cream}\hrule width 3.5in height 0.4pt \color{black}\vspace*{5pt}} 
\setcounter{secnumdepth}{5}

\makeatletter 
\renewcommand\@biblabel[1]{#1}            
\renewcommand\@makefntext[1]%
{\noindent\makebox[0pt][r]{\@thefnmark\,}#1}
\makeatother 
\renewcommand{\figurename}{\small{Fig.}~}
\sectionfont{\sffamily\Large}
\subsectionfont{\normalsize}
\subsubsectionfont{\bf}
\setstretch{1.125} 
\setlength{\skip\footins}{0.8cm}
\setlength{\footnotesep}{0.25cm}
\setlength{\jot}{10pt}
\titlespacing*{\section}{0pt}{4pt}{4pt}
\titlespacing*{\subsection}{0pt}{15pt}{1pt}

\fancyfoot{}
\fancyfoot[LO,RE]{\vspace{-7.1pt}\includegraphics[height=9pt]{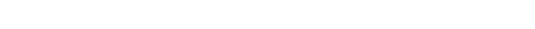}}
\fancyfoot[CO]{\vspace{-7.1pt}\hspace{13.2cm}\includegraphics{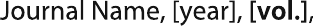}}
\fancyfoot[CE]{\vspace{-7.2pt}\hspace{-14.2cm}\includegraphics{RF}}
\fancyfoot[RO]{\footnotesize{\sffamily{1--\pageref{LastPage} ~\textbar  \hspace{2pt}\thepage}}}
\fancyfoot[LE]{\footnotesize{\sffamily{\thepage~\textbar\hspace{3.45cm} 1--\pageref{LastPage}}}}
\fancyhead{}
\renewcommand{\headrulewidth}{0pt} 
\renewcommand{\footrulewidth}{0pt}
\setlength{\arrayrulewidth}{1pt}
\setlength{\columnsep}{6.5mm}
\setlength\bibsep{1pt}

\makeatletter 
\newlength{\figrulesep} 
\setlength{\figrulesep}{0.5\textfloatsep} 

\newcommand{\topfigrule}{\vspace*{-1pt}%
\noindent{\color{cream}\rule[-\figrulesep]{\columnwidth}{1.5pt}} }

\newcommand{\botfigrule}{\vspace*{-2pt}%
\noindent{\color{cream}\rule[\figrulesep]{\columnwidth}{1.5pt}} }

\newcommand{\dblfigrule}{\vspace*{-1pt}%
\noindent{\color{cream}\rule[-\figrulesep]{\textwidth}{1.5pt}} }

\makeatother

\twocolumn[
  \begin{@twocolumnfalse}
{\includegraphics[height=1pt]{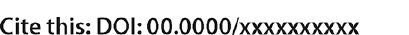}\hfill\raisebox{0pt}[0pt][0pt]{\includegraphics[height=55pt]{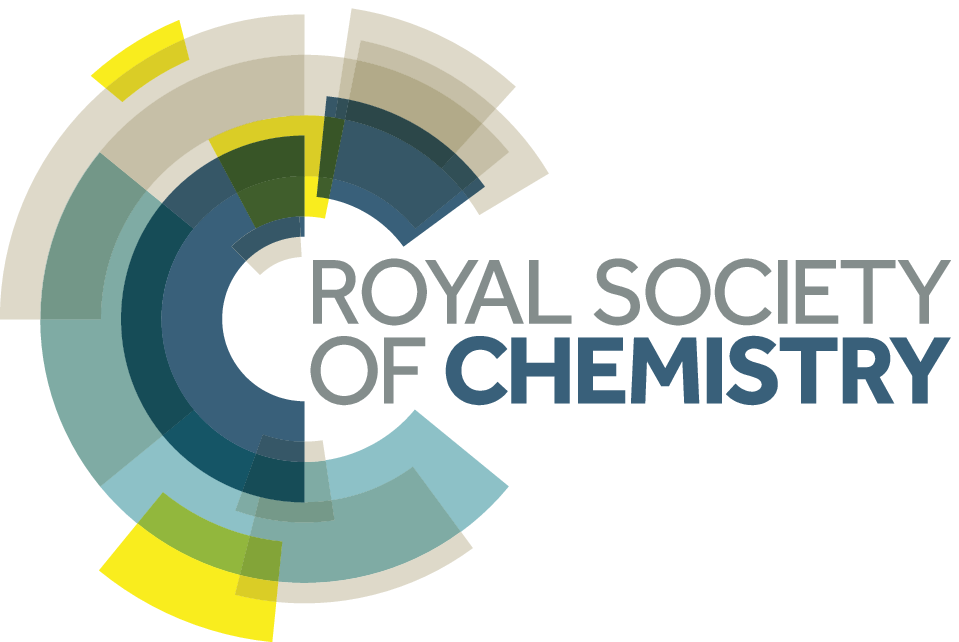}}\\[1ex]
\includegraphics[width=18.5cm]{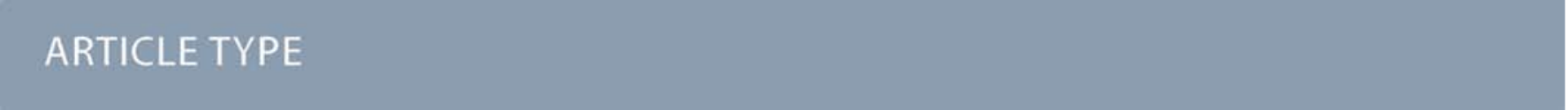}}\par
\vspace{1em}
\sffamily
\begin{tabular}{m{4.5cm} p{13.5cm} }

\includegraphics{DOI} & \noindent\LARGE{Anionic nickel and nitrogen effects in the chiral antiferromagnetic antiperovskite Mn$_3$NiN} \\
\vspace{0.3cm} & \vspace{0.3cm} \\

 & \noindent\large{E. Triana-Ramírez$^{a,b}$, W. Ibarra-Hernandez$^{c}$\,\orcidlink{0000-0002-5045-4575} and A. C. Garcia-Castro$^{a,*}$\orcidlink{0000-0003-3379-4495}} \\

\includegraphics{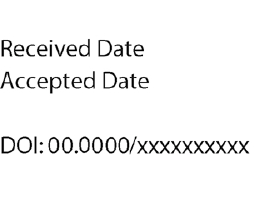} & \noindent\normalsize{Magnetic antiperovskites, holding chiral noncollinear antiferromagnetic ordering, have shown remarkable properties that cover from negative thermal expansion to anomalous Hall effect. Nevertheless, details on the electronic structure related to the oxidation states and the octahedral center's site effect are still scarce. 
Here, we show a theoretical study, based on first-principles calculations in the framework of the density-functional theory, DFT, on the electronic details associated with the nitrogen site effect into the structural, electronic, magnetic, and topological degrees of freedom. 
Thus, we show that the nitrogen vacancy increases the values of the anomalous Hall conductivity and retains the chiral $\Gamma_{4g}$ antiferromagnetic ordering.
Moreover, we reveal, based on the Bader charges and the electronic structure analysis, the negative and positive oxidation states in the Ni and Mn sites, respectively. 
The latter is in agreement with the expected $A_3^{\alpha+}B^{\beta-}X^{\delta-}$ oxidation states to satisfy the charge neutrality in the antiperovskites, but rare for transition metals. 
Finally, we extrapolate our findings on the oxidation states to several Mn$_3B$N compounds showing that the antiperovskite structure is an ideal platform to encounter negative oxidation states in metals sitting at the corner $B$-site.} 

\end{tabular}

 \end{@twocolumnfalse} \vspace{0.6cm}

  ]

\renewcommand*\rmdefault{bch}\normalfont\upshape
\rmfamily
\section*{}
\vspace{-1cm}


\footnotetext{\textit{$^{a}$~School of Physics, Universidad Industrial de Santander, Carrera 27 Calle 09, 680002, Bucaramanga, Colombia}}
\footnotetext{\textit{$^{b}$~Centro de Investigaci\'on y Estudios Avanzados del IPN - $\textsc{cinvestav}$-Quer\'etaro,  MX-76230, Quer\'etaro, M\'exico. }}
\footnotetext{\textit{$^{c}$~Facultad de Ingenier\'ia, Benem\'erita Universidad Aut\'onoma de Puebla, Apartado Postal J-39, Puebla, Pue. 72570, M\'exico.}}
\footnotetext{\textit{*E-mail: acgarcia@uis.edu.co}}

\section{Introduction:}
\label{sec1}
In the modern quest for novel and exciting phenomena in new materials, antiperovskite, or inversed-perovskite with the formulae $A_3BX$, has stood out as an astonishing type of materials showing large anomalous Hall conductivity \cite{Gurung2019,Liu2018,Tsai2020,PhysRevB.106.195113}, superconductivity \cite{Oudah2016, Oudah2019, Wang2013, Okoye_2003}, good performance for new batteries \cite{doi:10.1021/ja305709z, D1TA03680G}, tunable hybrid-improper ferroelectricity \cite{Garcia-Castro2019}, tangible magnetocaloric effects \cite{Lewis_2006}, and large spin-lattice coupling \cite{FLOREZGOMEZ2022169813, doi:10.1021/acsami.8b03112, PhysRevB.78.184414, PhysRevB.96.024451, Gomonaj1989}.
All the latter are just a few examples that can be mentioned among the vast functionalities offered by these materials. 
Interestingly, in antiperovskites \cite{Wang2019, Garcia-Castro2020} the electrostatic balance and the oxidation site occupation are apparently reversed with respect to the known perovskites,  $ABX_3$ \cite{Krivovichev+2008+109+113}. 
Most of the mentioned phenomena in the antiperovskites owe most of their properties to the reversed occupation of their anionic and cationic sites forming the reversed $XA_6$ octahedra.
This subtle change in the coordination and atomic occupations gives rise to, for example, the triangular geometric coordination of the magnetic sites that in turn, induces a strong magnetic frustration converging into chiral noncollinear antiferromagnetic orderings \cite{Fruchart1978}. Another example is the topologically related properties in the Sr$_3$SnO (Sr$_3$PbO) with bands crossing at the Fermi level between the Sr:4$d$ and Sn:5$p$ (Pb:6$p$) electronic states \cite{PhysRevB.100.245145,PhysRevMaterials.3.124203}. Here, the Sn and Pb atomic sites hold formal negative oxidation states confirmed by Mössbauer \cite{PhysRevB.100.245145} and X-ray photoemission spectroscopy, XPS \cite{PhysRevMaterials.3.124203}.
Interestingly, the negative oxidation states in metals have attracted considerable attention due to the possible new physics that may unveil \cite{Ellis2006, CHAN201343}.
Despite these reports, few metallic species are known in the literature. Some examples of negative oxidation states in metals have been demonstrated in compounds such as CsAu \cite{B708844M} and Na-Au binary compounds \cite{PhysRevB.84.014117, Sarmiento_P_rez_2013}. 
In the latter compounds, gold's oxidation state is Au$^{1-}$ achieving a full 6$s^2$5$d^{10}$ electronic occupation in the outer shell. 
Other examples are the Pt's negative oxidation at Ba-Pt systems \cite{doi:10.1021/ja0401186,https://doi.org/10.1002/anie.200462055} and dimethylformamide's surface \cite{doi:10.1021/jp068879d,doi:10.1021/ja071483a} as well as the negative oxidation state in Zn at octahedrally coordinated Zn$_2M_4$ ($M$ = Li and Na) \cite{https://doi.org/10.1002/cphc.201901051}. 
Moreover, multiple molecular compounds have shown metals in their structures with formal negative oxidation states \cite{Ellis2006, CHAN201343}.
As such, antiperovskites appear to be potential candidates to explore possible negative oxidation states in metals,  and their induced properties, due to the expected stoichiometric relationship $A_3^{\alpha+}B^{\beta-}X^{\delta-}$ in contrast to $A^{\alpha+}B^{\beta+}X_3^{\delta-}$ to hold the charge neutrality. 
Then, when going from the perovskite to the antiperovskite, the $B$-site switches from $B^{\beta+}$ to  $B^{\beta-}$. 
This is the case of the SrSnO$_3$ and Sr$_3$SnO where the Sn oxidation state goes from $4+$ to $4-$ \cite{PhysRevB.100.245145} to maintain the charge neutrality in both compounds, while the Sr and O sites keep the $2+$ and $2-$ oxidation, respectively. 
These findings are also in agreement with other gold-based antiperovskites oxides Cs$_3$AuO and Rb$_3$AuO \cite{doi:10.1021/ja00152a016}.
Interestingly as observed in perovskites, the anionic vacancies, such as oxygen deficiency in perovskite oxides, \cite{PhysRevX.3.021010,PhysRevB.93.045405} could alter the electronic structure by inducing an electronic reconstruction that directly affects the presented oxidation states of the atomic species \cite{LSMO-2028}.
These vacancies can be present despite the advances in growth techniques nitrogen vacancies are expected to be formed, as in the Mn$_3$PtN$_x$ case \cite{doi:10.1142/S0217979218503149}. 
Therefore, exploring the effect of the nitrogen deficiency in the Mn$_3B$N antiperovskites, and particularly in the Mn$_3$NiN prototype, is essential in order to unveil its influence on the structural and electronic degrees of freedom that may affect the magnetic response and anomalous Hall conductivity.
Additionally, the N-site is strongly correlated with the oxidation states at the Mn- and Ni-sites in the Mn$_3$NiN antiperovskite.
In this paper, we study from first-principles calculations, the electronic structure of the chiral antiferromagnetic antiperovskite Mn$_3$NiN. 
Thus, we performed a detailed study of this antiperovskite's electronic structure and explored the formal charges of the atomic species coupled with the understanding of the N-site effect in the physics beneath the electronic structure of this compound. 
This antiperovskite stands as a prototype in this family of materials and we pay special attention to the influence of the nitrogen vacancy in the topological features, such as the anomalous Hall conductivity.
In Section \ref{secII} we present all the computational details and the theoretical approaches used for the development of this work. This section is followed by the presentation of the obtained results and the consequent analysis, in Section \ref{secIII}. 
Finally, we draw our conclusions, in Section \ref{conclusions}, and highlight some perspectives associated with our findings around the oxidation states and the nitrogen's effect in the Mn$_3$NiN antiperovskite. Furthermore, we extrapolate these analyzes and include our results in several Mn$_3B$N antiperovskites.

\section{Theoretical and computational details:}
\label{secII}
We performed first-principles calculations within the density-functional theory (DFT) \cite{PhysRev.136.B864,PhysRev.140.A1133} approach by using the \textsc{vasp} code (version 5.4.4) \cite{Kresse1996,Kresse1999}. 
The projected-augmented waves, PAW \cite{Blochl1994, PhysRevB.59.1758} scheme was used to represent the valence and core electrons.
The electronic configurations considered in the pseudo-potentials as valence electrons are Mn: (3$p^6$3$d^5$4$s^2$, version 02Aug2007), Ni: (3$p^6$3$d^8$4$s^2$, version 06Sep2000), and N: (2$s^2$2$p^5$, version 08Apr2002). 
The exchange-correlation, $E_{xc}$, was represented within the generalized gradient approximation, GGA in the PBEsol parametrization \cite{Perdew2008}, and the $E_{xc}$ of the $d$-electrons was corrected through the DFT$+U$ approximation within the Liechtenstein formalism \cite{Liechtenstein1995}. 
We used a Coulomb on-site value of $U$ = 2.0 eV parameter. The latter optimized to reproduce the experimentally observed  lattice parameter. 
Also, a metaGGA formalism \cite{PhysRevB.84.035117}, within the SCAN implementation \cite{PhysRevLett.115.036402}, and the hybrid based-functional, HSE06 \cite{doi:10.1063/1.2404663} were adopted to correlate with the Hubbard correction within the PBEsol$+U$ calculations.
The periodic solution of the crystal was represented by using Bloch states with a Monkhorst-Pack \cite{PhysRevB.13.5188} \emph{k}-point mesh of 13$\times$13$\times$13 and 600 eV energy cut-off to give forces convergence of less than 0.001 eV$\cdot$\r{A}$^{-1}$ and an error in the energy less than 0.5 meV.  
The spin-orbit coupling (SOC) was included to consider noncollinear magnetic configurations \cite{Hobbs2000}.  
The phonon calculations were performed within the finite-differences methodology \cite{PhysRevLett.48.406, PhysRevB.34.5065} and analyzed through the \textsc{phonopy} interface \cite{phonopy,PhysRevLett.78.4063}. The latter calculations were performed in the 2$\times$2$\times$2 supercell to properly map the lattice dynamics at the zone-boundary. For these calculations in the supercell, the $k$-mesh was then set to 6$\times$6$\times$6 and the noncollinear magnetic orderings were also considered.
To evaluate the anomalous Hall conductivity, and the changes in the Berry curvature, we have used the Wannier functions methodology for which the wannierization was performed with the \textsc{Wannier90} code \cite{MOSTOFI20142309, Pizzi_2020} and post-processed with the \textsc{WannierBerri} package \cite{wannierberri}. Here, $s$, $p$, and $d$ orbitals were considered in the Mn and Ni cases, while $s$ and $p$ were considered at the N site. Additionally, a 3.0 eV window was used around the Fermi level for the wannierization.
Bader charges were evaluated by the methodology developed by G. Henkelman \emph{et al.}   \cite{HENKELMAN2006354}.
Finally, the atomic structure figures were elaborated with the \textsc{vesta} code \cite{vesta}.

\section{Results and discussion:}
\label{secIII}

\begin{figure*}[t]
 \centering
 \includegraphics[width=15.0cm,keepaspectratio=true]{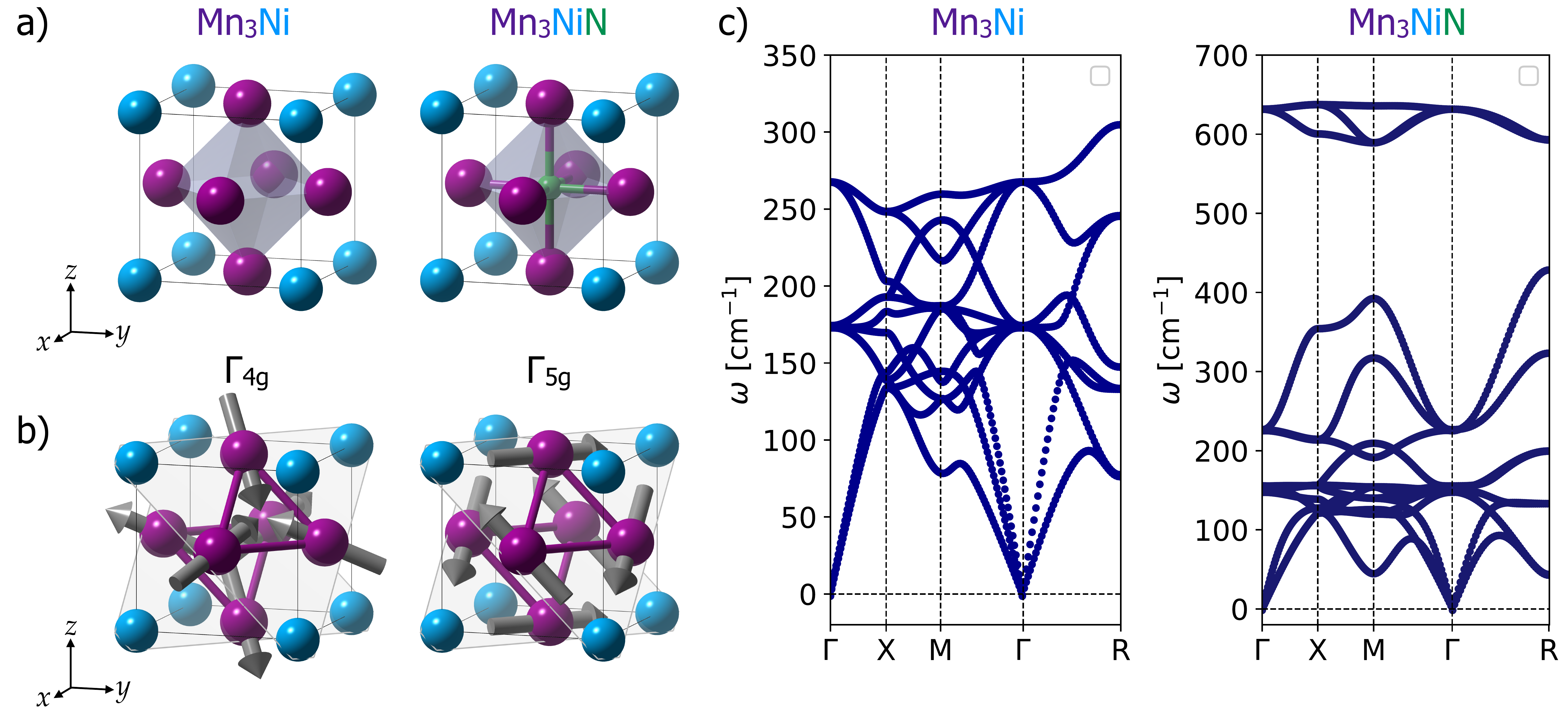}
 \caption{(Color online) (a) Mn$_3$Ni and Mn$_3$NiN $Pm\bar{3}m$ structures where the N-site octahedral center is shown in the latter and absent in the former. In (b) are shown the chiral antiferromagnetic noncollinear $\Gamma_{4g}$ and $\Gamma_{5g}$ orderings. Here, the magnetic moments per Mn atom are shown as grey arrows by notation. In (c) are presented the full phonon-dispersion curves obtained for the Mn$_3$NiN, as well as the nitrogen-deficient Mn$_3$Ni antiperovskites. The latter were computed with $U$ = 2.0 eV. In both cases, we consider the $\Gamma_{4g}$ chiral antiferromagnetic ordering ground state.}
 \label{F1}
\end{figure*} 

In what follows, we start by describing the electronic properties related to the N-site effect in the Mn$_3$NiN antiperovskite. In Fig. \ref{F1} are shown the Mn$_3$Ni and Mn$_3$NiN cubic $Pm\bar{3}m$ (SG. 221) antiperovskites as well as the symmetry allowed noncollinear chiral antiferromagnetic $\Gamma_{4g}$ and $\Gamma_{5g}$ orderings. 
Thus, in the nitrogen deficiency antiperovskite, the $\Gamma_{4g}$ ordering is more stable over the  $\Gamma_{5g}$ explained in terms of the MAE energy $\Delta E = E_{\Gamma_{4g}} - E_{\Gamma_{5g}}$ = $-$0.58 meV$\cdot$f.u.$^{-1}$.
Thus, as in the Mn$_3$NiN case, the magnetic ground state in the Mn$_3$Ni is the chiral $\Gamma_{4g}$ antiferromagnetic ordering that allows the anomalous Hall effect \cite{PhysRevLett.112.017205}, as will be discussed further. 
Therefore, it is worth recalling that all the calculations contained in this work were performed considering the spin-orbit coupling and the noncollinear antiferromagnetic states for Mn$_3$NiN, as well as for Mn$_3$Ni antiperovskites.
After fully relaxing the Mn$_3$NiN and Mn$_3$Ni the obtained lattice parameters are $a$ = 3.889 \r{A} and $a$ = 3.707 \r{A} respectively. It can be noted that, in the Mn$_3$NiN case, the lattice parameter is in good agreement with the experimentally reported value of $a$ = 3.886 \r{A} \cite{doi:10.1063/1.4822023}. 
In the Mn$_3$Ni, although there is no experimentally reported parameter, as the exchange-correlation correction was also considered in the Mn:3$d$ orbitals, it can be expected a close value to the one reported here by us. 
When comparing the lattice parameter, it can be observed that the inclusion of the N-site in the octahedral center induces a tangible lattice expansion, equivalent to 0.182 \r{A}. 
However, the symmetry space groups remain the same being $Pm\bar{3}m$ (SG. 221) without considering the magnetic ground state and $R\bar{3}m'$ (MSG. 166.101) once the chiral noncollinear antiferromagnetic ground state is accounted into the symmetry operations. 
Then, only changes in the volume and electronic structure were found. Moreover, both Mn$_3$NiN and Mn$_3$Ni antiperovskite are fully dynamically stable in the cubic configuration under the noncollinear $\Gamma_{4g}$ antiferromagnetic ordering, see Fig. \ref{F1}. As it can be observed, the high-frequency modes are absent in Mn$_3$Ni in which case, these modes are nitrogen driven.
For instance, the antiperovskite Mn$_3B$N structure can be viewed as magnetic Mn-based kagome lattices with $B$-sites embedded into them and separated by nitrogen layers.

\begin{figure*}[]
 \centering
 \includegraphics[width=18.3cm,keepaspectratio=true]{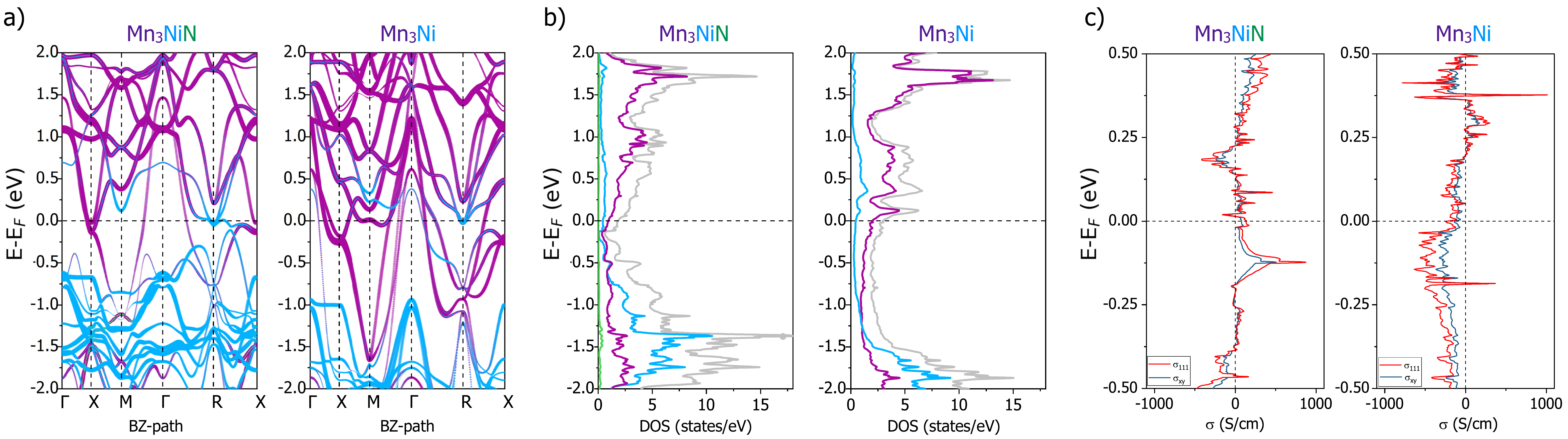}
 \caption{(Color online) (a) Atomically projected band structure, and (b) atomically projected density of states, DOS. Here, the Mn, Ni, and N states are denoted in violet, blue, and green colors respectively. Additionally, in (b) the total DOS is denoted in grey color. (c) Anomalous Hall conductivity, $\sigma_{xy}$ and $\sigma_{111}$ components, computed at the $\Gamma_{4g}$ orderings in the Mn$_3$NiN and Mn$_3$Ni.}
 \label{F2}
\end{figure*}

In Fig. \ref{F2} we present the entire electronic characterization for the Mn$_3$NiN and Mn$_3$Ni antiperovskites. Here, the full orbitally-projected band electronic structure is presented, in Fig. \ref{F2}(a), as well as the local density of states, in Fig. \ref{F2}(b), and the computed anomalous Hall conductivity for the $\sigma_{xy}$, and $\sigma_{111}$ terms, in Fig. \ref{F2}(c). 
At first glance, we can observe from Fig. \ref{F2}(a) that there is a substantial reduction of the available states close to the Fermi energy, here located at $E_F$ = 0.0 eV by notation, once the N-site is introduced in the antiperovskite.
It can be appreciated that the major contribution at and above the Fermi level is associated with the Mn:3$d$ orbitals in both cases.
As for the Ni states, those appear to be located well below $E$ =  $-$0.5 eV and are quite localized around $-$1.5 eV as in an insulator case, see Fig. \ref{F2}(b). 
Nevertheless, a small contribution from the Ni states can be observed above the Fermi level. The latter is expected because the antiperovskite structure can be understood, as commented before, as (111) Mn-based kagome planes with Ni sites embedded and separated by the N-sites.
Importantly, as the N-site is located at the octahedral center, the Mn$_3$NiN and the Mn$_3$Ni hold the same crystallographic and magnetic symmetry. Thus, only modifications in the electronic structure are observed, but the AHC tensor is kept fixed. 
In this case, the anomalous Hall conductivity component, $\sigma_{xy}$, has been computed by the relationship: 

\begin{equation}\label{eq:ahc}
   \sigma_{xy}=-\frac{2\pi e^2}{h} \sum_n^{occ} \int_{BZ} \frac{d^3\bf{k}}{(2\pi)^3} f_n({\bf{k}}) \Omega_{n,xy} (\bf{k}),
\end{equation}

where $\Omega_{xy}(\bf{k})$=$\sum_n^{occ} f_n(\bf{k})$$\Omega_{n,xy}(\bf{k})$ is the summation of all the occupied $n$-bands and $f_n\bf(k)$ represents the Fermi distribution.
Moreover, the symmetry-allowed AHC components within the $\Gamma_{4g}$ ordering in the $R\bar{3}m'$ magnetic symmetry group are:

\begin{equation}\label{ahc:tensor}
\sigma_{R\bar{3}m'}=
\begin{pmatrix}
0 & \sigma_{xy} & -\sigma_{xy}\\
-\sigma_{xy} & 0 & \sigma_{xy} \\
\sigma_{xy} & -\sigma_{xy}& 0
\end{pmatrix}
\end{equation}
\\

The charge at the Ni-site is expected to be the same and only changes in the allowed electronic states in the proximity to the Fermi level might influence the anomalous Hall conductivity. 
Additionally, as the AHC is strongly dependent on the spin-orbit coupling strength \cite{RevModPhys.82.1539}, the absence or presence of the nitrogen octahedral center site has a negligible effect on $\sigma_{xy}$.
In Fig. \ref{F2}(c) we show the computed anomalous Hall conductivity for the $\sigma_{xy}$ in both compounds, as well as the $\sigma_{111}$ component in the magnetic kagome lattice at the (111) lattice plane. The $\sigma_{111}$ component is computed as $\sigma_{111} \equiv \frac{1}{\sqrt{3}}\left(\sigma_{xy}+\sigma_{yz}+\sigma_{zx}\right)$ and corresponds to the conductivity on the (111) kagome lattice.
We then found that in absence of the N-site $\sigma_{xy}$ = 139 S$\cdot$cm$^{-1}$ ($\sigma_{111}$ = 241 S$\cdot$cm$^{-1}$) whereas in the Mn$_3$NiN is $\sigma_{xy}$ = 78 S$\cdot$cm$^{-1}$ ($\sigma_{111}$ = 135 S$\cdot$cm$^{-1}$), both at the $E_F$ level,.
Our findings show a considerable increase of the $\sigma_{xy}$ in the nitrogen-deficient antiperovskite that can be correlated to the increase of the available electronic states close to Fermi.  
The latter enhancement of the $\sigma_{xy}$ component is in agreement with the electronic band structure, also shown and discussed before. 
Therefore, the N-site is directly influencing the $f_n(\bf{k})$ function into Eq. \ref{eq:ahc} modifying the $\sigma_{xy}$ value but keeping the symmetry operations.

\begin{table}[t]
\centering
\caption{Bader charges, in $e^-$ units, computed for the Mn, Ni, and N sites in the Mn$_3$NiN and Mn$_3$Ni considering the chiral antiferromagnetic $\Gamma_{4g}$ ordering. The latter values were extracted under several exchange-correlation representations. Additionally, we present the magnetic moment, per Mn atom, in each case.}
\begin{tabular}{c  c  c  c | c  c  c}
\hline
\hline
 XC$_{PBEsol}$ &  $Z_{Mn}$ &  $Z_{Ni}$ & $Z_N$   &  $m$ ($\mu_B$$\cdot$Mn$^{-1}$)  \rule[-1ex]{0pt}{3.5ex} \\
\hline
Mn$_3$NiN &  $+$0.907 &  $-$0.723 & $-$2.006   & 3.560 \rule[-1ex]{0pt}{3.5ex}  \\
Mn$_3$Ni &  $+$0.259 &  $-$0.704 & ---  &  3.583 \rule[-1ex]{0pt}{3.5ex} \\
\hline
 XC$_{SCAN}$ &  $Z_{Mn}$ &  $Z_{Ni}$ & $Z_N$   &  $m$ ($\mu_B$$\cdot$Mn$^{-1}$)  \rule[-1ex]{0pt}{3.5ex}   \\
\hline
Mn$_3$NiN &  $+$0.806 &  $-$0.745 & $-$1.957  & 3.418  \rule[-1ex]{0pt}{3.5ex}  \\
Mn$_3$Ni &  $+$0.262 &  $-$0.788 & ---  & 3.470 \rule[-1ex]{0pt}{3.5ex} \\
\hline
 XC$_{HSE06}$ &  $Z_{Mn}$ &  $Z_{Ni}$ & $Z_N$   &  $m$ ($\mu_B$$\cdot$Mn$^{-1}$)  \rule[-1ex]{0pt}{3.5ex}   \\
\hline
Mn$_3$NiN &  $+$0.973 &  $-$0.790 & $-$2.130 & 3.854  \rule[-1ex]{0pt}{3.5ex}  \\
Mn$_3$Ni &  $+$0.324 &  $-$0.718 & ---  & 3.883 \rule[-1ex]{0pt}{3.5ex} \\
\hline
\hline
\end{tabular}
\label{tab:1}
\end{table}

In regards to the electronic structure, formally, the oxidation states according to the IUPAC \cite{IUPAC,KarenMcArdleTakats+2016+831+839,KarenMcArdleTakats+2014+1017+1081}, quantifies the oxidation degree of an atom defined based-on the electron counting of such atomic species after the bonding is reached. 
Therefore, the oxidation number can be obtained by following a set of rules, as exposed by A. Walsh \emph{et al.} \cite{doi:10.1021/acs.jpclett.7b00809,Walsh2018} that, as mentioned before, can be ascribed to the electron counting.
Then, aiming to estimate the potential oxidation states, hold by each atomic component in the antiperovskite, we proceded by obtaining the charges around each site.
As in the Mn$_3$NiN and Mn$_3$Ni antiperovskites, the electronic structure is metallic, the Born effective charges, $Z^*$, are not accessible due to the ill-defined polarization in metals\footnote{The $Z^*$ tensor is defined as $Z_{\alpha\beta,\kappa}^*$ = $\frac{\partial P_{\beta}}{\partial \tau_{\alpha,\kappa}}|_{\mathcal{E}=0}$ where $\alpha$ and $\beta$ are the cartesian coordinates, $P_{\beta}$ is the polarization and $\tau$ are the atomic displacements of the $\kappa$ atom.}, and therefore, the Bader charges offer an alternative route to estimate the charges in the atomic species. 
In Table \ref{tab:1} are condensed the results related to the Bader charges computed in the Mn$_3$NiN and Mn$_3$Ni antiperovskites. These values were obtained for the PBEsol$+U$, SCAN, and HSE06 exchange-correlation approaches. 
We can observe that, independently of the exchange-correlation considerations, following the Jacob's ladder from the GGA$+U$ to the hybrid-functional approach \cite{doi:10.1063/1.1904565}, the computed charges are close to $+$0.9$e^-$, $-$0.7$e^-$, and $-$2.0$e^-$ for the Mn, Ni, and N sites, respectively, in the Mn$_3$NiN case. 
The previous charges are in contrast with the computed charges of $+$0.3$e^-$ and $-$0.7$e^-$ for Mn and Ni in the Mn$_3$Ni. 
These results are suggesting a good representation of the charges with the PBEsol$+U$ approach. Thus, the PBEsol$+U$ exchange-correlation is used for further analysis.
As expected, the Mn-sites hold, in both antiperovskite cases, a positive charge associated with a Mn$^{\alpha+}$ oxidation state. 
Meanwhile, the corner Ni-site shows a negative charge leading to a Ni$^{\beta-}$ oxidation state. 
In the nitrogen case, as expected the Bader charge is negative and it is related to the N$^{\delta-}$, ($\delta$ = 3) oxidation state expected in this anionic site.
Interestingly, the Mn's Bader charge is $+$0.259$e^-$ in the Mn$_3$Ni whereas is $+$0.907$e^-$ in the Mn$_3$NiN. 
This can be explained due to the charge localized in the nitrogen site when incorporated in the antiperovskite and that it is transferred from the manganese sites. 
Moreover, this result is in agreement with larger electronic states, close to the Fermi level, available in the Mn$_3$Ni in comparison to Mn$_3$NiN, and also explaining the AHC results.
Aiming to compare with other insulating antiperovskites, such as Ca$_3$SnO and Ca$_3$BiN, we have computed the Bader charges and found that $Z_{Ca}$ = $+$1.308$e^-$, $Z_{Sn}$ = $-$2.364$e^-$, and $Z_{O}$ = $-$1.558$e^-$ for the Ca$_3$SnO oxide, and  $Z_{Ca}$ = $+$1.333$e^-$,  $Z_{Bi}$ = $-$1.955$e^-$, $Z_{N}$ = $-$2.043$e^-$, for the Ca$_3$BiN nitride antiperovskite case.
As for Born effective charges, $Z^*$ accessible in these compounds, we observed that the diagonal terms are $Z^*_{Ca}$ = $+$2.388$e^-$, $Z^*_{Sn}$ = $-$3.023$e^-$, and $Z^*_{O}$ = $-$3.381$e^-$ in the Ca$_3$SnO oxide, and $Z^*_{Ca}$ = $+$2.380$e^-$,  $Z^*_{Bi}$ = $-$2.899$e^-$, $Z^*_{N}$ = $-$4.397$e^-$ for the Ca$_3$BiN nitride. 
The deviation of the Born effective charges, with respect to the nominal values ($Z_{Ca}$ = $+$2$e^-$, $Z_{Sn}$ = $-$4$e^-$,  $Z_{Bi}$ = $-$3$e^-$, $Z_{O}$ = $-$2$e^-$ and $Z_{N}$ = $-$3$e^-$,), can be explained in terms of the large polarizability of the Sn--O and Bi--N bondings widely observed and reported in ferroelectric perovskite oxides \cite{PhysRevB.58.6224, PhysRevB.51.6765}.
Despite the charge underestimation, shown by the Bader analysis, and overestimation, obtained with the Born effective charges, the latter results are in fair agreement with the expected oxidation states of $A_3^{2+}B^{4-}$O$^{2-}$ and $A_3^{2+}B^{3-}$N$^{3-}$ compounds, respectively.
As such, these findings are consistent with the experimentally measured, by Mössbauer spectroscopy and X-ray photoemission spectroscopy, XPS, oxidation states of the atomic constituents in the Sr$_3$SnO and Sr$_3$PbO antiperovskites \cite{PhysRevB.100.245145, PhysRevMaterials.3.124203}. 
In such compounds, the oxidation state was associated with Sn$^{4-}$ and Pb$^{4-}$ states based on the experimental results. 
Additionally, these results on the oxidation states are also in agreement with the calculations in other antiperovskite insulators such as Ba$_3$SiO and Ba$_3$SiO/Ba$_3$GeO ferroelectric superlattices in which, the $Z^*$ values are $+$2.396$e^-$, $-$4.720$e^-$, $-$4.594$e^-$, and $-$2.801$e^-$ for the Ba, Si, Ge, and O sites, respectively \cite{Garcia-Castro2019,GARCIACASTRO2020109126}.

To contrast the obtained oxidation states in the antiperovskite Mn$_3$NiN, we defined a hypothetical perovskite compound as NiMnN$_3$ by inverting the Mn and N sites. 
Thus, the Mn occupies the octahedral center whereas the N sites form the octahedra, $i.e.$ MnN$_6$. 
Here, the Ni sites remain in the cell's corner site. 
In such a compound, we have fully relaxed the structural and electronic degrees of freedom and found a metallic behavior with a tangible magnetic response in which, the Mn holds a $m$ = 2.501 $\mu_B$$\cdot$Mn$^{-1}$ and the Ni is $m$ = $-$1.080 $\mu_B$$\cdot$Ni$^{-1}$. 
After extracting the Bader charges we obtained values of  $Z_{Mn}$ = $+$1.981$e^-$, $Z_{Ni}$ = $+$0.888$e^-$, $Z_{N}$ = $-$0.957$e^-$. 
As expected, the $A^+B^+X_3^-$ is kept as Ni$^{\alpha+}$Mn$^{\beta+}$N$_3^{\delta-}$.
It is worth mentioning that the Bader charges seem to underestimate the computed charge in the atomic sites, as observed, for example in Ca$_3$BiN possibly due to the partitioning methodology and exchange-correlation approach  \cite{doi:10.1021/acs.jctc.0c00440}.
Nevertheless, it can be concluded that as perovskite and antiperovskite structures are considered, the Ni-site oxidation state is reversed from positive, in the former, to negative in the latter.

\begin{table}[t]
\centering
\caption{Computed Bader charges, in $e^-$ units, for the Mn, $B$-sites, and N sites in the Mn$_3B$N  within the chiral antiferromagnetic $\Gamma_{4g}$ ordering. Here, we also present the magnetic moment, per Mn atom in each case as well as the electronic configuration for each $B$-site with the outer valence electrons in neutral state. In the Mn and Ni cases, the outer electrons are [Ar]4s$^2$3$d^5$ and [He]2$s^2$2$p^3$, respectively.}
\begin{tabular}{c  c  c  c | c}
\hline
\hline
Mn$_3B$N &  $Z_{Mn}$ &  $Z_{B}$ & $Z_N$   & $B$-site   \rule[-1ex]{0pt}{3.5ex} \\
\hline
Mn$_3$NiN &  $+$0.907 &  $-$0.723 & $-$2.006   & Ni:[Ar]4$s^2$3$d^8$ \rule[-1ex]{0pt}{3.5ex}  \\ 
Mn$_3$PdN &  $+$0.884 &  $-$1.023 & $-$1.629   & Pd:[K]5$s^0$4$d^{10}$  \rule[-1ex]{0pt}{3.5ex}  \\ 
Mn$_3$PtN &  $+$0.982 &  $-$1.351 & $-$1.596   & Pt:[Xe]6$s^1$5$d^9$  \rule[-1ex]{0pt}{3.5ex}  \\ 
Mn$_3$IrN &  $+$0.946 &  $-$1.273 & $-$1.566   & Ir:[Xe]6$s^2$5$d^7$  \rule[-1ex]{0pt}{3.5ex}  \\
\hline
\hline
Mn$_3$NiN &  $+$0.907 &  $-$0.723 & $-$2.006   & Ni:[Ar]4$s^2$3$d^8$  \rule[-1ex]{0pt}{3.5ex}  \\ 
Mn$_3$CuN &  $+$0.754 &  $-$0.601 & $-$1.661   & Cu:[Ar]4$s^1$3$d^{10}$  \rule[-1ex]{0pt}{3.5ex}  \\
Mn$_3$ZnN &  $+$0.682 &  $-$0.391 & $-$1.656   & Zn:[Ar]4$s^2$3$d^{10}$ \rule[-1ex]{0pt}{3.5ex}  \\ 
Mn$_3$GaN &  $+$0.593 &  $-$0.136 & $-$1.643   & Ga:[Ar]4$s^2$3$d^{10}$4$p^1$  \rule[-1ex]{0pt}{3.5ex}  \\ 
Mn$_3$SnN &  $+$0.661 &  $-$0.355 & $-$1.630   & Sn:[Kr]5$s^2$4$d^{10}$5$p^2$  \rule[-1ex]{0pt}{3.5ex}  \\ 
\hline
\hline
\end{tabular}
\label{tab:2}
\end{table}

Moving forward, we applied our analysis to several members of the Mn$_3B$N family in order to extrapolate our findings. 
In Table \ref{tab:2} we present our calculations of the Bader charges across several reported Mn-based antiperovskites. In all the cases, the chiral antiferromagnetic $\Gamma_{4g}$ ordering was considered as the magnetic ground state in the calculations.
As expected, the N-sites remain negative with values between $-$2.0$e^-$ to $-$1.6$e^-$ whereas the Mn-sites vary from $+$0.9$e^-$ to $+$0.6$e^-$.
Thus, we followed the trend vertically along the periodic table from the 3$d$ in Ni to 5$d$ in Pt and horizontally from Ni to Ga.
We found an increase of the charge at the $B$-site, from Ni to Pt, suggesting an increase of the negative oxidation state, for example, $\beta$ $\sim$ 1$-$ in Ni to $\beta$ $\sim$ 2$-$ in Pt.
In the Mn$_3$IrN case, the Bader charge is close to the value observed for the Pt site. Interestingly, the received charge could be possibly located at the open $s$- and $d$-orbitals.
On the contrary, the charge decreases from Ni to Ga possibly due to the tendency, in this case, to keep the outer electronic shells closed.
As for the magnetic moment value, per Mn site, remains in values between 3.5 $\mu_B$$\cdot$Mn$^{-1}$ to 3.8 $\mu_B$$\cdot$Mn$^{-1}$. 
In contrast, we observed that the charge at the $B$-sites decreases when going from Ni to Ga. This can be explained in terms of the closing electronic shell limiting the space for acquired charge and therefore, diminishing the possible negative oxidation state.

Although we are aware of the underestimated charge obtained through the Bader charges approach, and fluctuations in the values of the charges could be expected due to the partitioning method employed \cite{doi:10.1021/acs.jctc.0c00440, Walsh2018}, our findings consistently suggest negative oxidation states in metals when located at the antiperovskite's corner $B$-site.
Finally, it is worth noticing that as the spin-orbit coupling, SOC, increases with the oxidation state \cite{https://doi.org/10.1002/jcc.21011}, the $B$-site's oxidation is paramount to understanding its contribution to the anomalous Hall conductivity. 
Thus, in the case of the Mn$_3B$N compounds, the SOC possibly decreases due to the negative oxidation state in the $B$-site when compared with perovskite compounds.

\section{Conclusions:}
\label{conclusions}
In this paper, we have studied the electronic structure of the Mn$_3$NiN and Mn$_3$Ni magnetically chiral noncollinear antiferromagnetic antiperovskites by utilizing first-principles calculations. 
We found that the N-site expands the cell when it is located at the center of the octahedral. Nonetheless, due to the center position, the symmetry operations and expected properties are conserved. 
We observed a tangible increase of the available electronic states close to the Fermi level that favors the conductivity, as in the case of the anomalous Hall effect in which, $\sigma_{xy}$ = 139 S$\cdot$cm$^{-1}$ ($\sigma_{111}$ = 241 S$\cdot$cm$^{-1}$) in absence of the N-site in contrast to $\sigma_{xy}$ = 78 S$\cdot$cm$^{-1}$ ($\sigma_{111}$ = 135 S$\cdot$cm$^{-1}$) in the Mn$_3$NiN counterpart.  
Our findings suggest that the nitrogen inclusion in the Mn$_3$NiN system enhances a positive oxidation state, possibly $\sim$1$+$, in the Mn whereas, and more interestingly, the Ni sites hold a negative, potentially $\sim$1$-$, oxidation state. This behavior is observed although the overall electronic structure remains metallic.
Finally, our findings also suggest that several transition metals may exhibit negative oxidation states when located at the $B$-site in the Mn$_3B$N antiperovskite. 
We thus hope that our result will motivate further studies in antiperovskite structures that might be ideal candidates to further investigate the negative oxidation states in metals.

\section*{Author contributions}
All of the authors were involved in the preparation and development of the manuscript.
Moreover, all of the authors read and approved the final manuscript.

\section*{Conflict of interest}
The authors declare no personnel or financial conflict of interests with any person or organization.

\section*{Acknowledgments:}
\label{acknowledgements}
The calculations presented in this paper were carried out using the Grid UIS-2 experimental testbed, being developed under the Universidad Industrial de Santander (SC3-UIS) High Performance and Scientific Computing Centre, development action with support from UIS Vicerrectoría de Investigación y Extension (VIE-UIS) and several UIS research groups as well as other funding resources.
Additionally, we acknowledge the computational support extended to us by Laboratorio de Supercomputo del Sureste de M\'exico (LNS), Benemérita Universidad Autónoma de Puebla, BUAP, for performing heavy theoretical calculations.
A. C. Garcia-Castro acknowledge the grant No. 2677 entitled “Quiralidad y Ordenamiento Magnético en Sistemas Cristalinos: Estudio Teórico desde Primeros Principios” supported by the VIE – UIS.



\balance


\bibliography{library} 
\bibliographystyle{rsc} 

\end{document}